\documentclass[sigconf]{acmart}

\usepackage{booktabs} 
\usepackage{amsmath}
\usepackage{relsize}
\usepackage[ruled,vlined,linesnumbered,noend]{algorithm2e}
\usepackage[noend]{algpseudocode}
\usepackage{url}
\usepackage[para]{footmisc}
\setcopyright{none}

\acmDOI{10.475/123_4}

\acmISBN{123-4567-24-567/08/06}

\acmConference[MLG'18]{$14^{th}$ International Workshop on Mining and Learning with Graphs}{August 2018}{London, United Kingdom}
\acmYear{2018}
\copyrightyear{2018}

\renewcommand\footnotetextcopyrightpermission[1]{}
\newcommand{\pattern}{pattern}
\newcommand{\mainAlgo}{SIAS-Miner-Enum}
\newcommand{\toERModel}{transformToER}
\newcommand{\pnrMiner}{P-N-RMiner}

\newcommand{\minimalDescription}{DL-Optimise}
\newcommand{\bestDesc}{\text{bestDesc}}
\newcommand{\cand}{\text{Cand}}
\newcommand{\pruneUseless}{pruneUseless}
\newcommand{\pruneLowerBounded}{pruneLowerBounded}
\newcommand{\csea}{CSEA}
\newcommand{\ic}{\ensuremath{\text{IC}}}
\newcommand{\dl}{\ensuremath{\text{DL}}}
\newcommand{\si}{\ensuremath{\text{SI}}}
\newcommand{\Prob}{\mbox{Pr}}
\newcommand{\Dom}{\mbox{Dom}}
\newcommand{\hA}{\hat{A}}
\newcommand{\ha}{\hat{a}}
\newcommand{\hc}{\hat{c}}

\newcommand{\errors}{\mbox{exc}}

\newtheorem{property}{Property}

\begin{document}
\title{Mining Subjectively Interesting Attributed Subgraphs}
\subtitle{Work-in-progress paper}

\author{Anes Bendimerad}
\authornote{The first two authors contributed equally to the paper.}
\affiliation{%
  \institution{Univ Lyon, INSA, CNRS UMR 5205}
  \streetaddress{20 av. A. Einstein}
  \city{F-69621 France}
}
\email{aabendim@liris.cnrs.fr}

\author{Ahmad Mel}
\authornotemark[1]
\affiliation{%
  \institution{IDLab, Ghent University}
  \streetaddress{Technologiepark-Zwijnaarde 19}
  \city{Ghent, Belgium}
}
\email{ahmad.mel@ugent.be}

\author{Jefrey Lijffijt}
\affiliation{%
  \institution{IDLab, Ghent University}
  \streetaddress{Technologiepark-Zwijnaarde 19}
  \city{Ghent, Belgium}
}
\email{jefrey.lijffijt@ugent.be}

\author{Marc Plantevit}
\affiliation{%
   \institution{Univ Lyon, UCBL, CNRS UMR 5205}
  \streetaddress{20 av. A. Einstein}
  \city{F-69622 France}
}
\email{marc.plantevit@liris.cnrs.fr}

\author{C\'eline Robardet}
\affiliation{%
   \institution{Univ Lyon, INSA, CNRS UMR 5205}
  \streetaddress{20 av. A. Einstein}
  \city{F-69621 France}
}
\email{celine.robardet@liris.cnrs.fr}

\author{Tijl De Bie}
\affiliation{%
  \institution{IDLab, Ghent University}
  \streetaddress{Technologiepark-Zwijnaarde 19}
  \city{Ghent, Belgium}
}
\email{tijl.debie@ugent.be}

\renewcommand{\shortauthors}{A. Bendimerad et al.}
\newcommand{\modif}[1]{{\color{blue}#1}}

\begin{abstract}
Community detection in graphs, data clustering, and local pattern mining
are three mature fields of data mining and machine learning.
In recent years, attributed subgraph mining is emerging as a new
powerful data mining task in the intersection of these areas.
Given a graph and a set of attributes for each vertex,
attributed subgraph mining aims to find cohesive subgraphs
for which (a subset of) the attribute values has exceptional values in some sense.
While research on this task can borrow from the three abovementioned fields,
the principled integration of graph and attribute data poses two challenges:
the definition of a pattern language that is intuitive and lends itself to efficient search strategies,
and the formalization of the interestingness of such patterns.
We propose an integrated solution to both of these challenges.
The proposed pattern language improves upon prior work in being both highly flexible and intuitive.
We show how an effective and principled algorithm can enumerate patterns of this language.
The proposed approach for quantifying interestingness of patterns of this language
is rooted in information theory,
and is able to account for prior knowledge on the data.
Prior work typically quantifies interestingness based on the cohesion of the subgraph
and for the exceptionality of its attributes separately,
combining these in a parameterized trade-off.
Instead, in our proposal this trade-off is implicitly handled in a principled, parameter-free manner.
Extensive empirical results confirm the proposed pattern syntax is intuitive,
and the interestingness measure aligns well with actual subjective interestingness.
\end{abstract}

\keywords{Graphs, Networks, Subjective Interestingness, Subgraph Mining}

%
%



\maketitle

\section{Introduction}\label{sec:introduction}


The availability of network data has surged both due to the success of social
media and ground-breaking discoveries in experimental sciences. Consequently,
graph mining is one of the most studied tasks for the data mining community.
The value of graphs stems from the presence of meaningful
relationships among the data objects (the vertices). These can be explored by
approaches as different as graph embeddings\cite{DBLP:journals/corr/abs-1709-07604}---which map the nodes of a graph into a low
dimensional space while preserving the local and global graph structure as well
as possible---, community detection\cite{fortunato2010community}---the
discovery of groups of vertices that somehow `belong together'---, or subgraph
mining---the identification of informative subgraphs.

Besides the relational structure, graphs may carry information in the form of
attribute-value pairs on vertices and/or edges. Such graphs are called
attributed graphs\cite{DBLP:conf/sdm/MoserCRE09,DBLP:journals/tkde/PradoPRB13,DBLP:journals/pvldb/SilvaMZ12}. We focus here on graphs with attribute-value pairs on vertices (vertex-attributed graphs). Mining interesting subgraphs in attributed graphs is challenging, both conceptually and computationally. A prominent problem is defining the interestingness of a subgraph. Desirable properties of subgraph would be that it is cohesive (the attribute values of the vertices are similar) and that the vertices form an easy to describe pattern in the graph (e.g., vertices should be close to each other). A specific form of subgraph mining is to look for exceptional subgraphs, i.e., subgraphs whose attribute values are cohesive within the subgraph but exceptional in the full graph.
%
%
%

Few works in this direction exist. For example Atzmueller et al.\cite{AtzmuellerDM16} study mining communities (densely connected subgraphs) that can also be described well in terms of attribute values, while Bendimerad et al.~\cite{bendimerad2017mining} look for exceptional subgraphs that are connected. Various quality measures are used in the first work and the second relies on Weighted Relative Accuracy. We introduce a new generically applicable interestingness measure for exceptional subgraph patterns based on Information Theory which is more flexible and can incorporate prior knowledge about the graph to steer the scoring of subgraph patterns. Besides, while previous works use certain hard constraints to arrive at subgraphs that are somehow interpretable, we integrate the interpretability into the interestingness measure. Hence, the trade-off between informativeness and interpretability can be made in a principled manner.

More specifically, we consider the problem to identify informative subgraphs that
can be concisely described. The informativeness of a subgraph depends on the
number of vertices it covers (more is better) and how surprising the statistics
(attribute values) of those vertices are. Surprise is important because showing
the user statistics they expect to see does not teach them anything. 
Vertices that are spread out over the graph, or that share no statistics cannot be summarized well, so the end-user cannot generalize over the structure of the vertices in a pattern
and hence the graph structure is effectively transmitted to a user without any compression. Here instead, we look for subgraphs that are both homogeneous and localized, hence possess shared properties 
which can be exploited to produce concise descriptions.

\begin{figure}[t]
\centering
\begin{tabular}{c}
\includegraphics[width=0.35\textwidth]{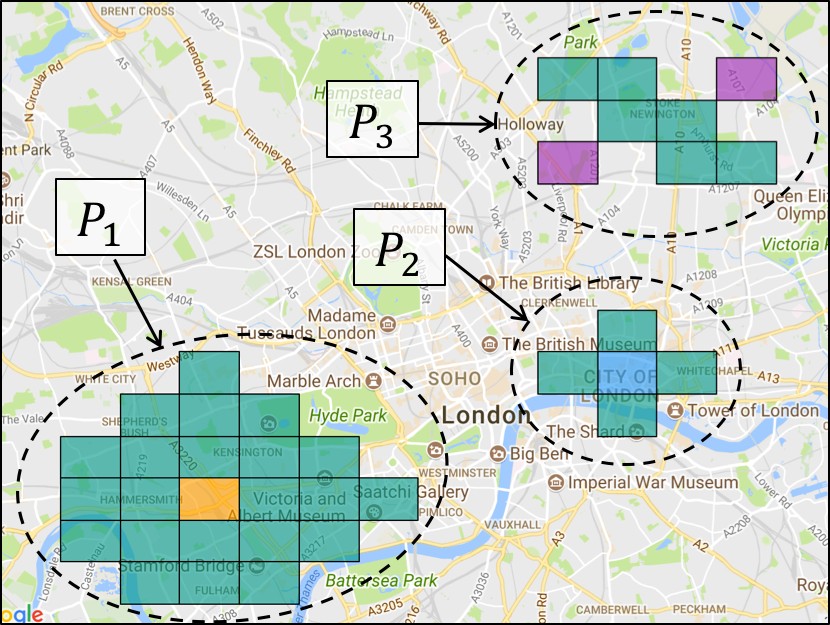} \\ 
    $P_{1}$: $\{$food$\} ^+ $ \\
    $P_{2}$: $\{$professional, nightlife, outdoors, college$\} ^+ $ \\
    $P_{3}$: $\{$nightlife, food$\} ^+ \{$college$\} ^- $ \\
\end{tabular}
   \caption{\label{fig:exampleGraphs} Example results on a graph
   based on Foursquare data covering the presence of various
   types of venues in London. $P_1$: around the orange block (west/south of Hyde Park) there are `surprisingly' many food establishments, except in the centre of that area (which is average). $P_2$: in the City several types of venues are consistently overrepresented. $P_3$: around Hackney there is a strip of blocks with lots of nightlife and food venues but limited educational venues.}
 \end{figure}

Fig.~\ref{fig:exampleGraphs} shows example patterns from our method, applied to
a setting where we want to explore the district structure of cities. The
patterns should be interpreted as follows: certain attributes have surprisingly
high/low values (marked $+$/$-$ respectively) in the given neighbourhoods as
compared to a background model.
We ensure the regions are localized by forcing them to have a
description of the form \textit{``all vertices that are within distance $d_1$ of vertex
$x$ AND within distance $d_2$ of vertex $y$ AND etc.''}; e.g.,
in pattern $P_3$ of Fig.~\ref{fig:exampleGraphs}, the covered areas (green blocks) are the
intersection of blocks within distance two of either purple block.

\textbf{Contributions. }  We present a pattern syntax for cohesive subgraphs with exceptional attributes (Sec.~\ref{sec:formulation}). We formalize their subjective interestingness in a principled manner using information theory, accounting for both the information they provide, as well as their interpretability (Sec.~\ref{sec:SI}). We study how to mine such subgraphs efficiently  (Sec.~\ref{sec:algorithms}). We provide a thorough empirical study on real data that evaluates (1) the
relevance of the subjective interestingness measure compared to state-of-the-art methods, and (2) the efficiency of the algorithms (Sec.~\ref{sec:experiments}). We discuss related work in Sec.~\ref{sec:relatedwork} and the conclusions are presented in Sec.~\ref{sec:conclusion}.






\section{Cohesive subgraphs with exceptional attributes\label{sec:formulation}}


Before formally introducing the pattern language we are interested in,
let us establish some notation.

{\bf Notation} An \emph{attributed graph} is  denoted $G=(V,E,\hA)$, where $V$ is a set of $n$
vertices, $E\subseteq V\times V$ is a set of $m$ edges, and $\hA$ is a
set of $p$ numerical attributes on vertices (formally, functions mapping a vertex onto an attribute value),
with $\ha(v)\in\Dom_a$ denoting the value of attribute $\ha\in \hA$ on $v\in V$.
We use hats in $\ha$ and $\hA$ to signify the empirical values of the attributes, 
whereas $a$ and $A$ denote (possibly random) variables over the same domains.
We also define the function $N_d(v)$ to denote the 
neighborhood of range $d$ of  a vertex $v$, i.e., the set of vertices whose geodesic distance to $v$ is at most $d$:
\[N_d(v)=\{u \in V \text{ } | \text{ } dist(v,u) \leq d \}.\]

{\bf Cohesive Subgraphs with Exceptional Attributes (\csea)} As described in the introduction,
we are interested in patterns that inform the user that a given set of attributes has exceptional values
throughout a set of vertices in the graph.

Thus, and more formally, a \emph{\csea{} pattern} is
defined as a tuple $(U,S)$, where $U \subseteq
V$ is a set of vertices in the graph,
and $S$ is a set of restrictions on the value domains of the attributes of $A$,
or more specifically, $S=\{ \lbrack k_a,\ell_a\rbrack\mid a\in A\}$.
A pattern $(U,S)$ is said to be contained in $G$ iff
\begin{equation}
\forall \lbrack k_a,\ell_a\rbrack\in S\ \mbox{and}\ \forall u\in
U,\: k_a\leq \hat{a}(u)\leq \ell_a. \label{eq:syntax}
\end{equation}


Informally speaking, a \csea{} pattern will be more informative if the ranges in $S$ are smaller,
as then it conveys more information to the data analyst.
We will make this more formal in Section~\ref{subsec:ic}.

At the same time, a \csea{} pattern $(U,S)$ will be more interesting if its description
is more concise in some \emph{natural easier-to-interpret definition}.
Thus, along with the pattern language,
we must also specify how a pattern from this language will be intuitively described.

To this end, we propose to describe the set of vertices $U$
as a neighborhood of a specified range from a given specified vertex,
or more generally as the intersection of a set of such neighborhoods.
For enhanced expressive power,
we additionally allow for the description to specify some exceptions on the above:
vertices that do fall within this (intersection of) neighborhood$(s)$,
but which are to be excluded from $U$. Exceptions are a detriment to the
interestingness of a pattern, but we can discount these naturally.

A premise of this paper is that this way of describing the set $U$ is intuitive for human analysts,
such that the length of the description of a pattern, as discussed in detail in Sec.~\ref{descLength},
is a good measure of the complexity to assimilate or understand it.
Our qualitative experiments in Sec.~\ref{sec:experiments} do indeed confirm this is the case.


\section{The subjective interestingness of a \csea{} pattern}\label{sec:SI}

The previous sections already hinted at the fact that we will formalize the interestingness of a \csea{} pattern $(U,S)$
by trading off its information content
with its description length.
Here we show how the FORSIED framework for formalizing subjective interestingness of patterns,
introduced in~\cite{DBLP:conf/kdd/Bie11,DBLP:journals/datamine/Bie11}, can be used for this purpose.

The information content depends on both $U$ and $S$.
It is larger when more vertices are involved, when
the intervals are narrower, and when they are more extreme.
We will henceforth denote the information content as $\ic(U,S)$.
The description length depends on
$U$ only (as the attribute ranges require a fixed description length),
and will be denoted as $\dl(U)$.
The subjective interestingness of a \csea{} pattern $(U,S)$ is then expressed as:
\[\si(U,S)=\frac{\ic(U,S)}{\dl(U)}.\]

One of the core capabilities of the FORSIED framework is that it quantifies the information content of a pattern
against a prior belief state about the data.
It rigorously models the fact that the more plausible the data is (subjectively) according to a user or (objectively) under a specified model,
the less information a user receives, and thus the smaller the information content ought to be.

This is achieved by
modeling the prior beliefs of the user
as the Maximum Entropy (MaxEnt) distribution subject
to any stated prior beliefs the user may hold about the data. This
distribution is referred to as the \emph{background distribution}. The
information content $\ic(U,S)$ of a \csea{} pattern $(U,S)$ is then formalized as minus the
logarithm of the probability that the pattern is present under the
background distribution (also called the self-information or surprisal)~\cite{cover1991entropy}:
\[\ic(U,S)=-\log(\Prob(U,S)).\]

In Sec.~\ref{subsec:ic}, we first discuss in greater detail
which prior beliefs could be appropriate for \csea{} patterns,
and how to infer the corresponding background distribution.
Then, in Sec.~\ref{descLength}, we discuss in detail
how the description length $\dl(U)$ can be computed.

\subsection{The information content of a \csea{} pattern}\label{subsec:ic}

{\bf Positive integers as attributes}
For concreteness, let us consider the situation where the attributes are positive integers $(a: V\:\rightarrow\mathbb{N}$, $\forall a\in A)$,
as will be our main focus throughout this paper.%
\footnote{The presented results can be extended relatively straighforwardly for other cases.}
For example, if the vertices are geographical regions (with edges connecting vertices of neighboring regions),
then the attributes could be counts of particular types of places in the region
(e.g.\ one attribute could be the number of shops).
It is clear that it is less informative to know that an attribute value is large
in a large region than it would be in a small region.
Similarly, a large value for an attribute that is generally large
is less informative than if it were generally small.
The above is only true, however, if the user knows (or believes) \emph{a priori}
at least approximately what these averages are for each attribute,
and what the `size' of each region is.
Such prior beliefs can be formalized as equality constraints
on the values of the attributes $A$ on all vertices,
or mathematically:
\begin{align*}
 \sum_{A} \Prob(A) \left(\sum_{a\in A} a(v) \right)= \sum_{\ha\in \hA} \ha(v),  & \quad \forall v \in V,  \\
 \sum_{A} \Prob(A) \left(\sum_{v\in V} a(v) \right)= \sum_{v\in V} \ha(v), &  \quad \forall a \in A.  
\end{align*}
The MaxEnt background distribution can then be found as the probability distribution $\Prob$ maximizing the entropy
$-\sum_{A} \Prob(A)\log\Prob(A)$,
subject to these constraints and the normalization $\sum_{A} \Prob(A) = 1$.

As shown in~\cite{DBLP:journals/datamine/Bie11},
the optimal solution of this optimization problem is a product of independent Geometric distributions,
one for each vertex attribute-value $a(v)$.
Each of these Geometric distributions is of
the form $\Prob(a(v)=z)=p_{av} \cdot (1-p_{av})^z $, $z \in
\mathbb{N}$, where $p_{av}$ is the success probability and it is
given by: $p_{av}=1-\exp(\lambda_a^r + \lambda_v^c)$, with
$\lambda_a^r $ and $ \lambda_v ^c$ the Lagrange multipliers corresponding to the two constraint types.
The optimal values of these multipliers can be found by solving the convex Lagrange dual optimization problem. 



Given these Geometric distributions for the attribute values under the background distribution,
we can now compute the probability of a pattern $(U,S)$ as follows:
%
\begin{align*}
\Prob(U,S) & = \prod_{v\in U}\prod_{\lbrack k_a,\ell_a\rbrack\in S} \Pr(a(v)\in\lbrack k_a,\ell_a\rbrack),\\
& = \prod_{v\in U}\prod_{\lbrack k_a,\ell_a\rbrack\in S} \left((1-p_{av})^{k_a}-(1-p_{av})^{\ell_a+1}\right).
\end{align*}
This can be used directly to compute the information content of a pattern
on given data, as the negative log of this probability.
However, the pattern syntax is not directly suited to be applied to count data,
when different vertices have strongly differing total counts.
The reason is that the interval of each attribute is the same across vertices,
which is desirable to keep the syntax understandable.
Yet, if neighboring regions have very different total counts, it is 
less likely to find any patterns, and, even if we do, end-users would still
need to know the total counts to interpret the patterns properly, as the same
interval is not equally surprising for each region.

{\bf $p$-values as attributes}
To address this problem, we propose to search for the patterns not on the counts themselves,
but rather on their \emph{significance} (i.e., $p$-value or tail probability),
computed with the background distribution as null hypothesis in a one-sided test.
More specifically, we define the quantities $\hc_a(v)$ as
\begin{align*}
\hc_a(v)&\triangleq \Prob(a(v)\geq\ha(v)),\\
&=(1-p_{av})^{\ha(v)},
\end{align*}
and use this instead of the original attributes $\ha(v)$.
This transformation of $\ha(v)$ to $\hc_a(v)$ can be regarded
as a principled normalization of the attribute values
to make them comparable across vertices.

To compute the \ic\ of a pattern with the transformed attributes $\hc_a$,
we must be able to evaluate the probability that $c_a(v)$
falls within a specified interval $\lbrack k_{c_a},\ell_{c_a}\rbrack$ under the background distribution for $a(v)$.
This is given by:
\begin{align*}
&\Prob(c_a(v)\in\lbrack k_{c_a},\ell_{c_a} \rbrack)=
\Prob\left((1-p_{av})^{a(v)}\in\lbrack k_{c_a},\ell_{c_a} \rbrack\right),\\
&=\Prob\left( a(v) \leq \frac{\log(k_{c_a})}{\log(1-p_{av})} \wedge a(v) \geq \frac{\log(\ell_{c_a})}{\log(1-p_{av})} \right),\\
&=(1-p_{av})^{\log_{1-p_{av}}(\ell_{c_a})}-(1-p_{av})^{\log_{1-p_{va}}(k_{c_a})+1},\\
&=\ell_{c_a}-(1-p_{av})\cdot k_{c_a},\\
&=\ell_{c_a}-k_{c_a}+p_{av}k_{c_a}.
\end{align*}

Thus, the \ic\ of a pattern on the transformed attributes $\hc$ can be calculated as:
%
  \begin{align}\label{eq:ic}
\ic(U,S)&=-\log(\Prob(U,S)),\nonumber\\
&=-\sum_{\lbrack k_{c_a},\ell_{c_a}\rbrack\in S} \sum_{v \in U} \log(\ell_{c_a} - k_{c_a} + p_{av}k_{c_a}).
  \end{align}

In this paper, we focus on intervals $\lbrack k_{c_a},\ell_{c_a}\rbrack$
where either $k_{c_a}=0$ (the minimal value) or $\ell_{c_a}=1$ (the maximal value).
Such intervals state that the values of an attribute are all significantly large\footnote{Note empty regions have tail probabilities $\hat{c}_{a}(v) = 1$ for any attribute and thus fall within any upper interval, but also $\ic = 0$ for any attribute of that region as both $l_{c_a} = 1$ and $p_{av} = 1$.} or significantly small respectively, for all vertices in $U$.
We argue such intervals are easiest to interpret.
The logarithmic terms in Eq.~(\ref{eq:ic}) then simplify to $\log(\ell_{c_a})$ and $\log(1-k_{c_a}+p_{av}k_{c_a})$ respectively.

\subsection{Description length}\label{descLength}


As mentioned above, we describe the vertex set $U$ in the pattern as
(the intersection of) a set of neighborhoods $N_d(v),\: v\in V$,
with a set of exceptions: vertices are in the intersection
but not part of $U$.
The length of such a description is the sum of the description lengths of the neighborhoods and the exceptions.
More formally, let us define the set of all neighborhoods $\mathcal{N}=\{N_d(v) \text{ } | \text{ } v \in V \wedge d
\in \llbracket 0, D \rrbracket \}$, (with $D$ a positive integer representing the radius of the neighborhood),
and let $\mathcal{N}(U)=\{N_d(v) \in \mathcal{N} \mid U \subseteq N_d(v) \}$ be the subset of neighborhoods that contain $U$.
The length of a description of $U$ as the intersection of all neighborhoods in a subset $X\subseteq\mathcal{N}(U)$,
along with the set of exceptions $\errors(X,U)\triangleq\cap_{N_d(v)\in X}N_d(v)\setminus U$,
is then quantified by the function
$f:2^{\mathcal{N}(U)}\times U\longrightarrow \mathbb{R}$:
\[f(X,U)=(|X|+1) \cdot \log(|\mathcal{N}|)+(|\errors(X,U)|+1) \cdot\log(|V|).\]
Indeed, the first term accounts for the description of the number of neighborhoods
($\log(|\mathcal{N}|)$, as there can be no more than $|\mathcal{N}|$ neighborhoods in $X$,
and for describing which neighborhoods are involved ($|X|\log(|\mathcal{N}|)$).
The second term accounts for the description of the number of exceptions ($\log(|V|)$),
and for describing the exceptions themselves ($|\errors(X,U)\log(|V|)$).

In general, there are multiple ways of describing a given set of nodes $U$, by using a various combinations of neighborhoods.
The best one is thus the one that minimizes $f$.
This finally leads us to the definition of the description length of a pattern as:
\[DL(U)=\min_{X \subseteq \mathcal{N}(U)} f(X,U).\]

\section{An enumeration approach to mining interesting \csea{} patterns\label{sec:algorithms}}
\mainAlgo{} mines interesting patterns using an enumerate-and-rank approach.
First, it enumerates all \csea{} patterns $(U,S)$ that are closed simultaneously
wrt. $U$, $S$, and the neighbourhood description. Second, it ranks
patterns according to their $\si$ values. An overview of the method is given in
Algorithm~\ref{algo3} and explained further below.

\begin{algorithm}[t]
    \caption{\mainAlgo($G=(V,E,\hat{A})$, $D$)\label{algo3}}
  \KwIn{$G$ the input graph, $D$ the maximum threshold of $d$ for used neighbourhoods $N_d(v)$.}
  \KwOut{$Result$, the set of \csea{s}, and $minDesc$, which stores the minimum description for each \pattern.}
  {\small
  // transformation to entity-relation model\\
  $\mathbb{D} \gets$ \toERModel($G$,$D$) \\
  // enumeration of the \pattern{s} \\ 
  Result $\gets $\pnrMiner($\mathbb{D}$)\\
  // computation of the minimum description for each found pattern \\
  minDesc $\gets \{ \}$ \\
  \For{$\langle(U,S),\mathcal{N}(U)\rangle \in $ Result}
      {\bestDesc$\gets \emptyset$\\
    \minimalDescription($U$, $\emptyset$,$\mathcal{N}(U)$, \bestDesc) \\
    minDesc[(U,S)] $\gets$ \bestDesc}
}
\end{algorithm}

\subsection{Pattern enumeration}\label{subsec:enumPatterns}

In the first step, we enumerate candidate tuples $(U,S)$ where the vertices $U$
can be concisely described as an intersection of neighborhoods
$X\subseteq\mathcal{N}(U)$. Disregarding the description for a moment, since the
pattern syntax is chosen such that each interval $[k_a,l_a] \in S$ should cover
every vertex $u \in U$ (Eq.~\ref{eq:syntax}), the \ic\ of a tuple $(U,S)$
increases monotonically by adding vertices to $U$ and intervals to $S$. Hence,
to decrease the number of candidate patterns and increase computational
efficiency we focus on \emph{closed} patterns, i.e., tuples $(U,S)$
where no
vertex can be added without enlarging intervals and where no interval
can be reduced without omitting vertices.

However, in Section~\ref{sec:formulation} we additionally argued that a pattern
$(U,S)$ where the vertices $u \in U$ are unrelated will be difficult to
understand and remember, which we expressed in the \dl. Since computing $\dl(U)$
is NP-Complete, it is not clear that enumeration of closed patterns wrt.\ the
\si\ can be done. Nonetheless, enumeration of all closed $(U,S)$ appears
wasteful because most sets $U$ will have a high \dl. What appears feasible is to
restrict enumeration of sets $U$ that are exactly
intersections of neighborhoods, i.e., a description without any exceptions%
\footnote{Notice that
  such a description does not
necessarily minimize the description length of a vertex set $U$.}.

While closed sets $(U,S)$ may be most efficiently enumerated by an itemset
mining algorithm, if we want $(U,S)$ to additionally be closed with respect to
intersections of neighbourhoods this yields a relational schema with two
relations: (1) vertices are connected to all intervals that cover their
attribute values, and (2) vertices are connected to every neighborhood that they
are contained in. A tool that would indeed enumerate precisely the required
closed patterns and no other is RMiner~\cite{DBLP:journals/datamine/SpyropoulouBB14}.
\begin{figure*}[htb]
\begin{center}
\includegraphics[width=.9\textwidth]{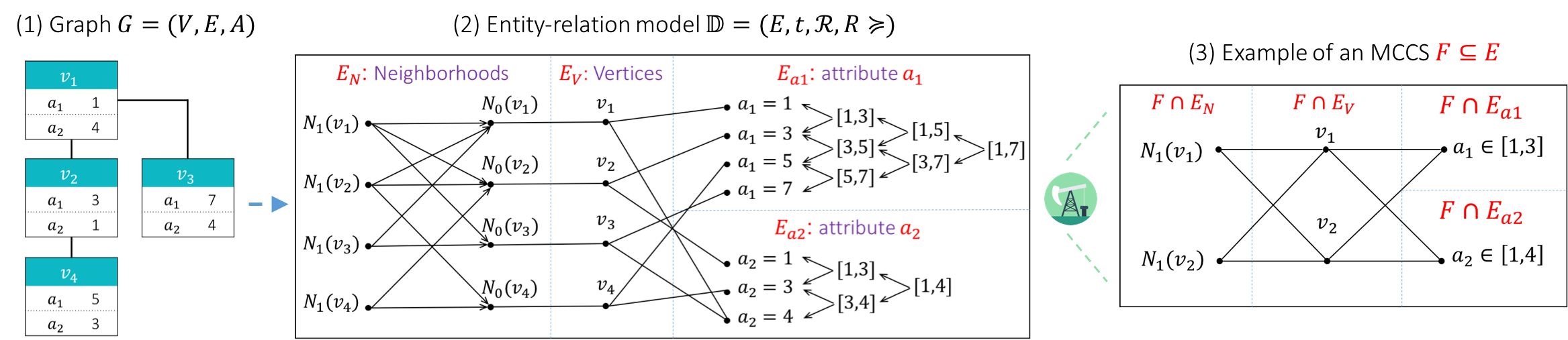}
\end{center} 
\caption{Transformation from (1) a graph structure to (2) an entity-relation
model with $D=1$, and (3) an example of a maximal complete connected subset
(MCCS) pattern from P-N-RMiner.}~\label{fig:toyExample}
\end{figure*}

More formally, the mapping from a graph to the required relational
format is depicted in Fig.~\ref{fig:toyExample}. Any vertex-attributed
graph $G$ can be mapped to an entity-relational model $\mathbb{D}$
through (1) creation of an entity type $E_v$ containing all vertices
in $G$, (2) creation of $|\hat{A}|$ entity types
$E_{a1}, E_{a2}, \ldots E_{a|\hat{A}|}$, one per attribute, and (3)
creation of an entity type $E_N$ containing all neighborhoods. 
Fig.~\ref{fig:toyExample} shows how intervals and
neighborhoods form a hierarchy. For neighborhoods, the relationship
holds that if a vertex $v_i$ is contained in a neighborhood
$N_k(v_j)$, then it is also contained in all neighborhoods with larger
hop-size:
$v_i \in N_k(v_j) \Rightarrow v_i \in N_l(v_j)\ \forall l \geq k$. A
similar statement holds for intervals.

P-N-RMiner~\cite{DBLP:journals/ijdsa/LijffijtSKB16} is an extension of RMiner
that exploits such hierarchies for efficiency. Firstly, fewer connections
(edges) are needed between the entity types hence there is a smaller memory
requirement. Secondly, there may be computational gains: P-N-RMiner is based on
fixpoint-enumeration~\cite{DBLP:journals/tcs/BoleyHPW10}, whose theory states
that efficient enumeration of closed sets is possible if and only if the problem
can be cast as a strongly accessible set system $(E,\mathcal{F})$.

The efficiency gain over plain enumeration comes from a closure operator, which
can skip non-closed candidate patterns. For P-N-RMiner, this closure
works outside-in, i.e., closed patterns are found by considering whether any
entity (vertex in the relational representation) can be added without reducing
the set of entities that are currently valid extensions to form a pattern under
the pattern syntax. Patterns in (P-N-)RMiner are called complete connected
subsets (CCSs) because all possible edges must exist and the vertices in the
relational representation must be connected, see Fig.~\ref{fig:toyExample},
right for an example. In the relational context, a pattern is called a maximal
CCS (MCCS) if no entity can be added, i.e., patterns we referred to as closed in
the discussion above. It is worth to notice that an MCCS provides, in addition
to a tuple $(U,S)$, the set $\mathcal{N}(U)$ of neighborhoods that contain $U$,
from which $DL(U)$ is computed.

Notice P-N-RMiner also ranks patterns based on interestingness under a
known-degree background model, but that is not useful in our setting as the \ic\
and the \dl\ are very different here. Hence, we only use it to enumerate all
candidates. The computational complexity is clearly exponential as the number of
outputs may be exponential in the size of the input, plus in this setting no
fixpoint-enumeration-based algorithm may have polynomial-time delay~\cite{DBLP:journals/ijdsa/LijffijtSKB16}. Scalability experiments are presented
in Sec.~\ref{sec:experiments}.

\subsection{Computing $DL(U)$}
\label{subsec:minimalDescCalculation}

The calculation of $DL(U)$ is NP-Complete and equivalent to Set Cover: it consists in finding the optimal cover of the set $\overline{U}$ based on unions of complements $\overline{N_i(v)}$ and exceptions $\{v \}$ such that $x \in \overline{U}$. Nevertheless, we propose a branch-and-bound approach that takes benefit from several optimisation techniques.

In order to find the optimal description of a pattern $(U,S)$, we
explore the search space $2 ^{\mathcal{N}(U)}$ with a branch-and-bound
approach described in Algorithm~\ref{algo4}. Let $X$ and $\cand$ be
subsets of $\mathcal{N}(U)$ that are respectively the current
enumerated description and the potential candidates that can be used
to describe $U$. Initially, \minimalDescription{} is called with
$X= \emptyset$ and $\cand=\mathcal{N}(U)$. In each call, a
neighbourhood $e \in \cand$ is chosen and used to recursively explore
two branches: one made of the descriptions that contain $e$ (by adding
$e$ to $X$), and the other one made of descriptions that do not
contain $e$ (by removing $e$ from $\cand$). Several pruning techniques are used in order to reduce the search
space and are detailed below. 
%

\begin{algorithm}[t]
  \caption{\minimalDescription ($U$, $X$, $\cand$, $\bestDesc$) \label{algo4}}
  \KwIn{$U$ the set of vertices to describe, $X$ the current enumerated  description, $\cand$ the set of candidates, $\bestDesc$ the current best description found.}
  \KwOut{$\bestDesc$ the best description found while exploring the current search sub-space.}
\If{$LB(X,U,\cand) < f(\bestDesc,U)$} 
    {\If{$\cand \neq \emptyset$} 
        {\pruneUseless($U$, $X$, $\cand$) \\
		\pruneLowerBounded($U$, $X$, $\cand$) \\		
		$e \gets \mbox{argmin}_{e' \in \cand} f(X \cup \{ e' \}, U)$ \\
		\minimalDescription($U$, $X \cup \{ e \}$, $\cand \setminus \{ e \}$, $\bestDesc$) \\
		\minimalDescription($U$, $X$, $\cand \setminus \{ e \}$, $bestDesc$) \\		
	}
	
	\uElseIf{$f(X,U) < f(\bestDesc,U)$ }
        {$\bestDesc \gets X$
	}
}
\end{algorithm}

{\bf Function LB (line 1)}
lower bounds the lengths of the descriptions that
can be generated in the subsequent recursive calls of
\minimalDescription. If $LB$ is higher or equal than the length of the
current best description of $U$ $f(\bestDesc,U)$, there is no need to
carry on the exploration of the search subspace as no further
description can improve $f$.
The principle of $LB$ is to evaluate the maximum reduction in exceptions that can be obtained when description $X$ is extended with
neighbourhoods of $Y$:
%
\begin{eqnarray}
  gain_Y(X,U)=\vert\errors(X,U)\vert - \vert\errors(X \cup Y, U)\vert,\mbox{ with }
  Y \subseteq \cand.
  \end{eqnarray}
This function can be rewritten using neighbourhood complements as
$gain_Y(X,U)=\vert \cup_{y \in Y} \left( \overline{y} \cap \errors(X,U) \right) \vert$
%
~\footnote
{$ =  |\errors(X,U)|-|\errors(X \cup Y, U)|  = |( \cap_{x \in X} x ) \setminus U| - |( \cap_{e \in X \cup Y} e ) \setminus U|$  \\
$ = |( \cap_{x \in X} x ) \cap \overline{U} | - |(( \cap_{x \in X} x ) \cap \overline{U} ) \cap ( \cap_{y \in Y} y )|$ \\
$ =  |( \cap_{x \in X} x ) \cap \overline{U}) \setminus ( \cap_{y \in Y} y ) |  =  |( \cap_{x \in X} x ) \setminus U) \cap \overline{ ( \cap_{y \in Y} y )  }|$ \\
$ =  |\errors(X,U) \cap  ( \cup_{y \in Y}  \overline{y} )  |  = | \cup_{y \in Y} ( \overline{y} \cap \errors(X,U) ) |$ \\
}.
We can obtain an upper bound of the gain function using the
ordered set $\{g_1,\ldots,g_{|\cand|} \}$ of
$\{ gain_{\{e\}}(X,U) \mid e \in \cand \}$ such that $g_i \geq g_j$ if
$i \leq j$:
\begin{property} \label{prop:sumgi}
  $gain_Y(X,U) \leq \sum_{i=1}^{|Y|} g_i$, for  $Y \subseteq \cand$.
\end{property}

\begin{proof}
Since the size of the union of sets is lower than the sum of the set sizes, we have
$gain_Y(X,U)  \leq \sum_{y \in Y} |\overline{y} \cap \errors(X,U)|
 \leq  \sum_{y \in Y} gain_{\{y\}}(X,U)  \leq \sum_{i=1}^{|Y|} g_i$.
\end{proof}
This is the foundation of the function $LB$ defined as 
\begin{eqnarray}
&  LB(X,U,\cand) = \min_{i \in \llbracket 0, |\cand| \rrbracket} \{ (|X|+i+1) \times \log(|\mathcal{N}|)  \nonumber \\
&   + \left(1+\max\left(0,|\errors(X,U)|-\sum_{j=1}^{i} g_i\right)\right) \times \log(|V|) \}
\end{eqnarray}

\begin{property}
  $f(X \cup Y, U) \geq LB(X,U,\cand)$, for all $Y \subseteq \cand$.
\end{property}
\begin{proof}
Based on Property~\ref{prop:sumgi}, we have
$|\errors(X \cup Y,U)| \geq \max ( 0,$ $|\errors(X,U)|- \sum_{i=1}^{|Y|} g_i )$.
This means that
$f(X \cup Y, U)  \geq  (|X|+|Y|+1) \times \log(|\mathcal{N}|)  + \left(1+\max\lbrace 0,|\errors(X,U)|-\sum_{j=1}^{|Y|} g_i\rbrace\right) \times log(|V|)$
and thus, $LB(X,U,\cand)  \leq  (|X|+|Y|+1) \times \log(|\mathcal{N}|)  
  + (1+\max\{ 0,|\errors(X,U)|$ $-\sum_{j=1}^{|Y|} g_i\}) \times \log(|V|)$  
  and it concludes the proof.
\end{proof}
In other terms,
in the recursive calls, a description length will never be lower than
$LB(X,U,\cand)$.

{\bf Function \pruneUseless{} line 3} removes candidate
elements that can not improve the description length, that is
candidates $e \in \cand$ for which $gain(\{e\},X,U) =0$. Such element
does not have the ability to reduce the number of exceptions in $X$. This
also implies that $e$ will not reduce the number of exceptions for
descriptions $X \cup Y$, with $Y \subseteq \cand$. Thus, such elements will not
decrease the description length of $X \cup Y$.

\begin{algorithm}[t]
  \caption{\pruneUseless ($U$, $X$, $\cand$) \label{algo:pruneUseless}}
  {\small
 $ \cand \gets \lbrace e \in \cand \mid \text{gain}(\{e\},X, U) > 0 \rbrace$
  }
\end{algorithm}


{\bf Function \pruneLowerBounded{} line 4}
%
removes a candidate $e \in \cand$ if there is a candidate
$e' \in \cand$ that is always better than $e$ for all descriptions produced in subsequent recursive calls.

\begin{property} \label{prop:pruneLBProof}
Let $e,e' \in \cand$ such that $\errors(X \cup \{e\},U) \subseteq \errors(X \cup \{e'\},U)$. Then, for all $Y \subseteq \cand \setminus \{e,e'\}$, we have $f(X \cup Y \cup \{e\}, U) \leq f(X \cup Y \cup \{e'\}, U) $
\end{property}

\begin{proof}
The set of exceptions in $X \cup Y \cup \{e\}$ is equal to $\errors(X \cup Y \cup \{e\}, U)= \errors(X \cup \{e\}, U) \cap \errors( Y, U)$. Since $\errors(X \cup \{e\},U) \subseteq \errors(X \cup \{e'\},U)$, then $\errors(X \cup Y \cup \{e\},U) \subseteq \errors(X \cup Y \cup \{e'\}, U)$. As $|X \cup Y \cup \{e\} |=|X \cup Y \cup \{e'\}|$, we can conclude that $f(X \cup Y \cup \{e\}, U) \leq f(X \cup Y \cup \{e'\}, U) $.
\end{proof}
\noindent Based on Property~\ref{prop:pruneLBProof},
\pruneLowerBounded{} removes elements $e' \in \cand$ such that
$\errors(X \cup \{e\},U) \subseteq \errors(X \cup \{e'\},U)$.  Notice
that even if an element $e''$ has been removed due to the lower bound
of $e'$, the procedure is still correct since $e''$ is lower
bound by $e$ by the transitivity of inclusion.

\begin{algorithm}[t]
  \caption{\pruneLowerBounded ($U$, $X$,$\cand$) \label{algo:pruneLB}}
  {\small
 $ \cand \gets \{ e_i \in \cand \mid \forall e_j  \in \cand \setminus \{ e_i \} : (\errors(X \cup \{e_j\},U) \not\subseteq \errors(X \cup \{e_i\},U))) 	\vee (\errors(X \cup \{e_j\},U) = \errors(X \cup \{e_i\},U)  \wedge i<j ) \}$
  }
\end{algorithm}

The last optimisation consists in choosing   $e\in\cand$
that minimises $f(X \cup \{ e \}, U)$ (line 5 of Algorithm~\ref{algo4}). This makes it possible
to quickly reach descriptions with low \dl , and subsequently provide effective pruning
when used in combination with $LB$.

\section{Experiments} \label{sec:experiments}

In this section, we report our experimental results. We start by describing the real-world dataset we used, as well as the questions we aim to answer. Then, we provide a thorough comparison with the state-of-the-art algorithm {Cenergetics} \citep{bendimerad2017mining}. Eventually, we provide a qualitative analysis that demonstrates the ability of our approach to achieve the desired goal.  For reproducibility purposes, the source code and the data are made available here.\footnote{https://goo.gl/2jvE8j}

\noindent {\bf Experimental setting.} Experiments are performed on the real-world dataset of the London graph.
The \emph{London graph} ($|V|=289 ,|E|=544 ,|\hA|=10$)
 is  based on the social network Foursquare\footnote{https://foursquare.com}. Each vertex represents a district in London, and edges
link adjacent districts.
Each attribute stands for
the number of places of a given type (e.g.\ outdoors, colleges,
  residences, restaurants, etc.) in each district.
  %
Considering all the numerical values of attributes is computationally  expensive and would lead to  redundant results,  we pre-process the graph so that for each attribute, the values $\hat{c}_a(v)$ are binned into five quantiles.


\noindent  {\bf Aims. } As stated in Section~\ref{sec:relatedwork}, there is no approach that supports the discovery of subjectively interesting attributed subgraphs in the literature. The closest method to \mainAlgo{} is {Cenergetics} \cite{bendimerad2017mining} that aims at discovering closed exceptional attributed subgraphs involving overrepresented and/or underrepresented attributes, and which mined the London graph used here in the experiments (and on similar graphs of other cities). It assesses exceptionality with the weighted relative accuracy (WRAcc) measure that accounts for margins but cannot account for other prior knowledge. The computational problem we tackle is more complex than {Cenergetics}, but how much is this overhead? Is it worth it in terms of pattern quality? This empirical study aims to answer to these questions.

\begin{figure*}[t]
\centering
 \begin{tabular}{cccc} \includegraphics[width=.23\textwidth]{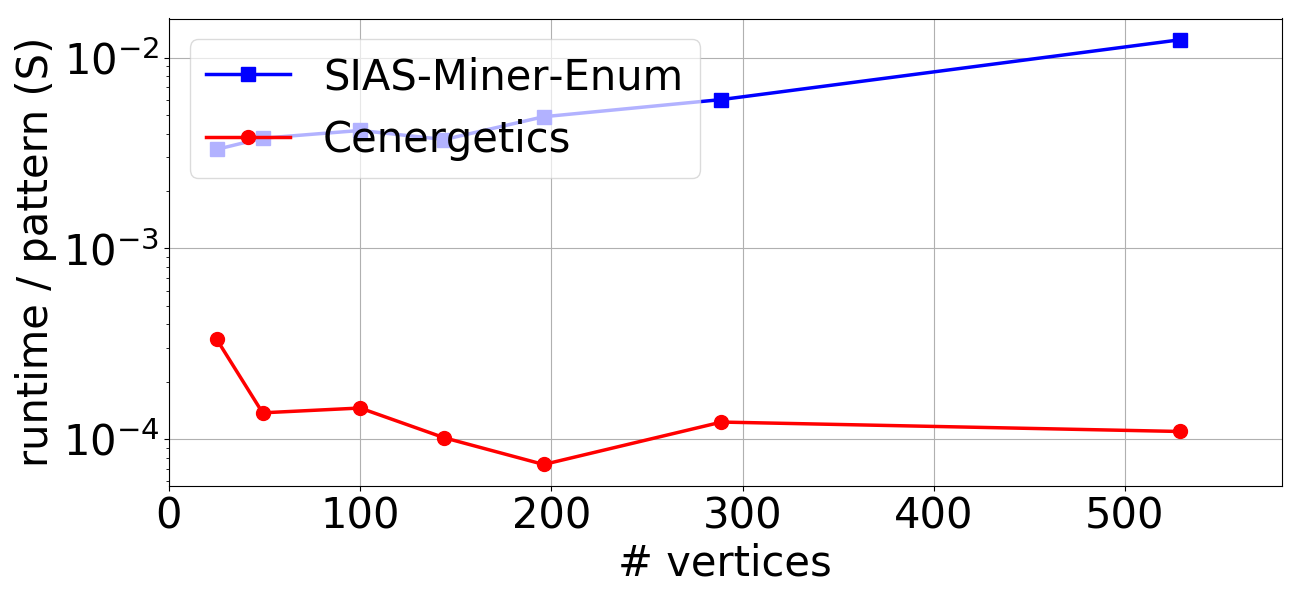}& \includegraphics[width=.23\textwidth]{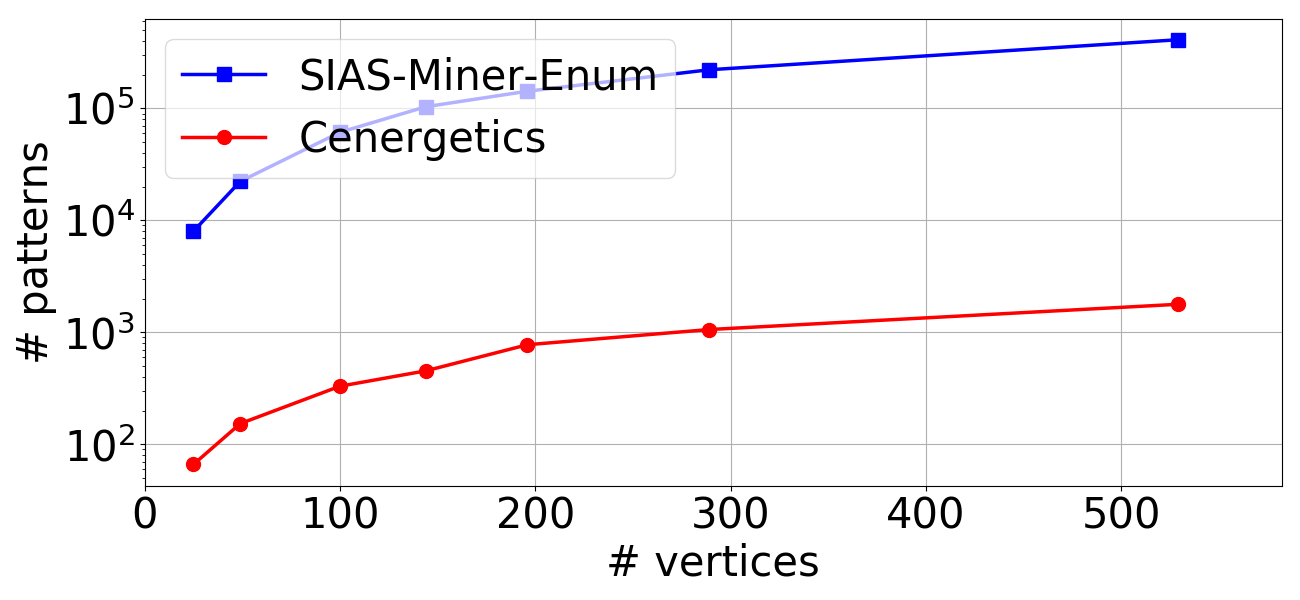} & 
 \includegraphics[width=.23\textwidth]{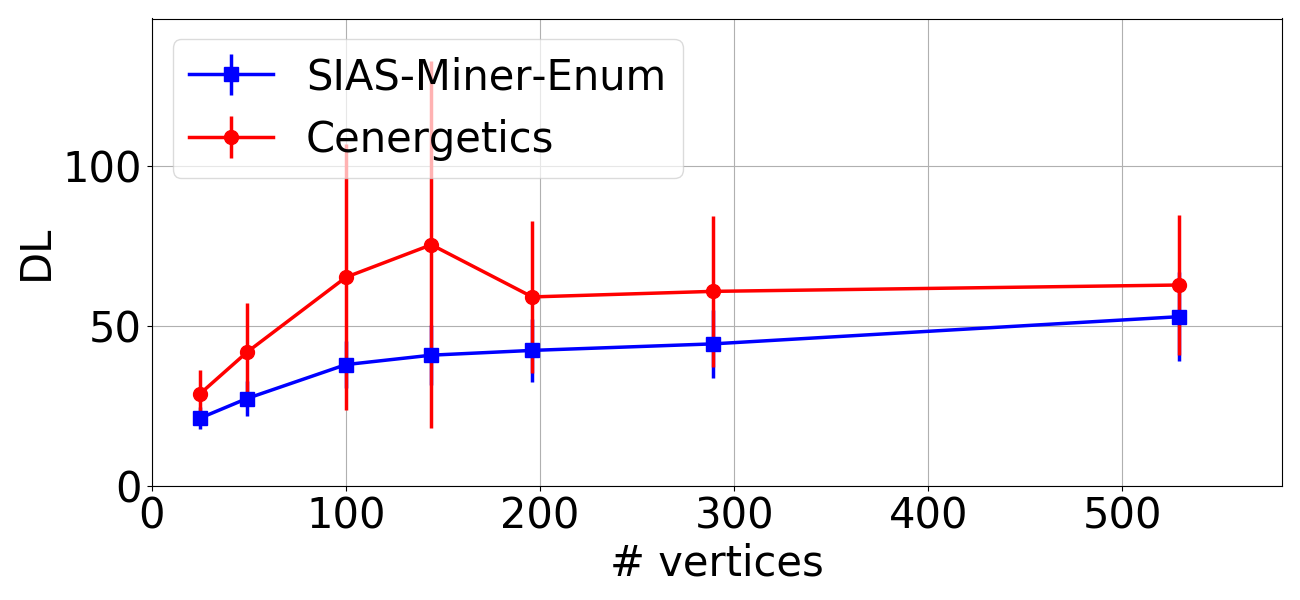} & 
  \includegraphics[width=.23\textwidth]{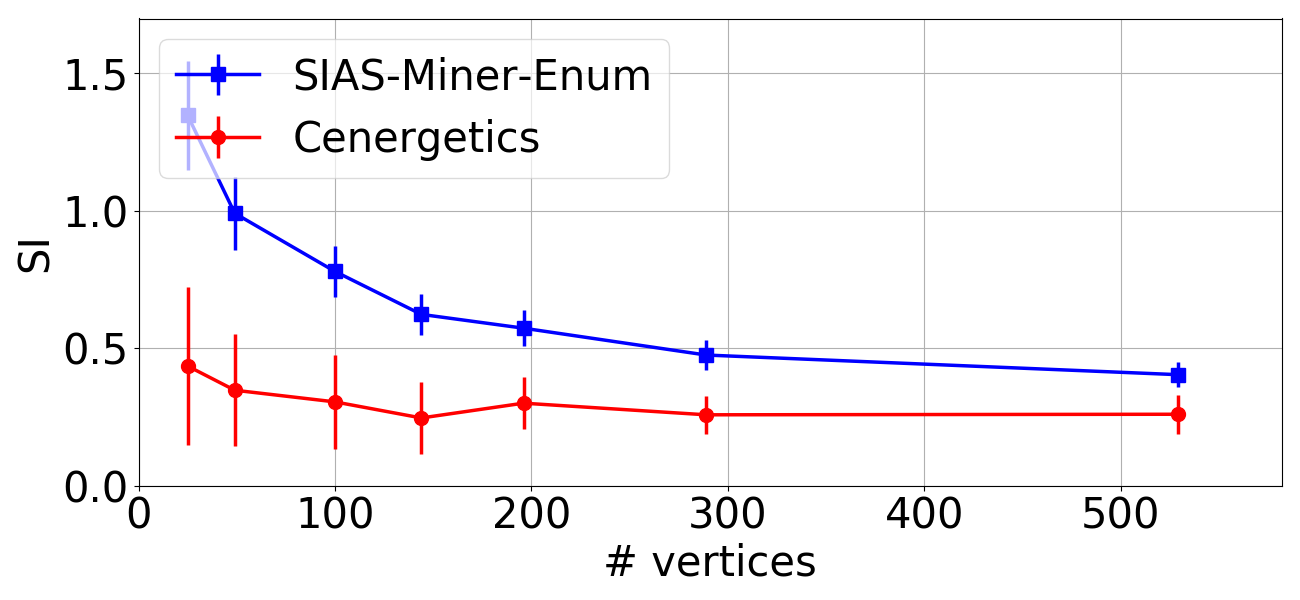}
 \\
    \includegraphics[width=.23\textwidth]{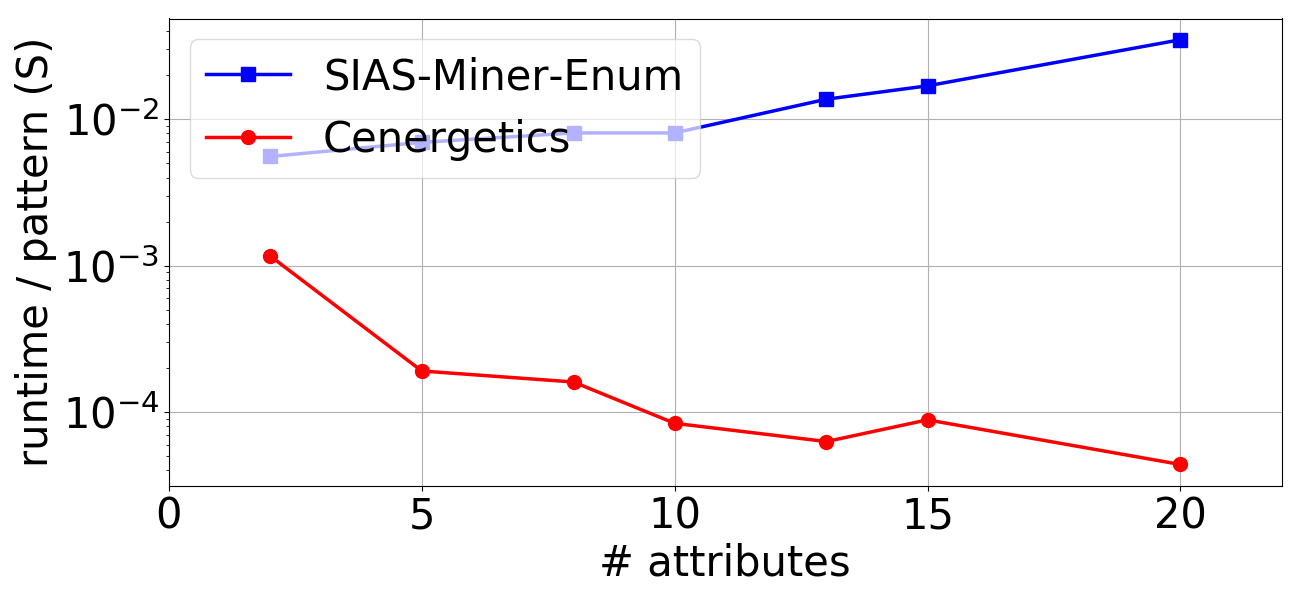}&
  \includegraphics[width=.23\textwidth]{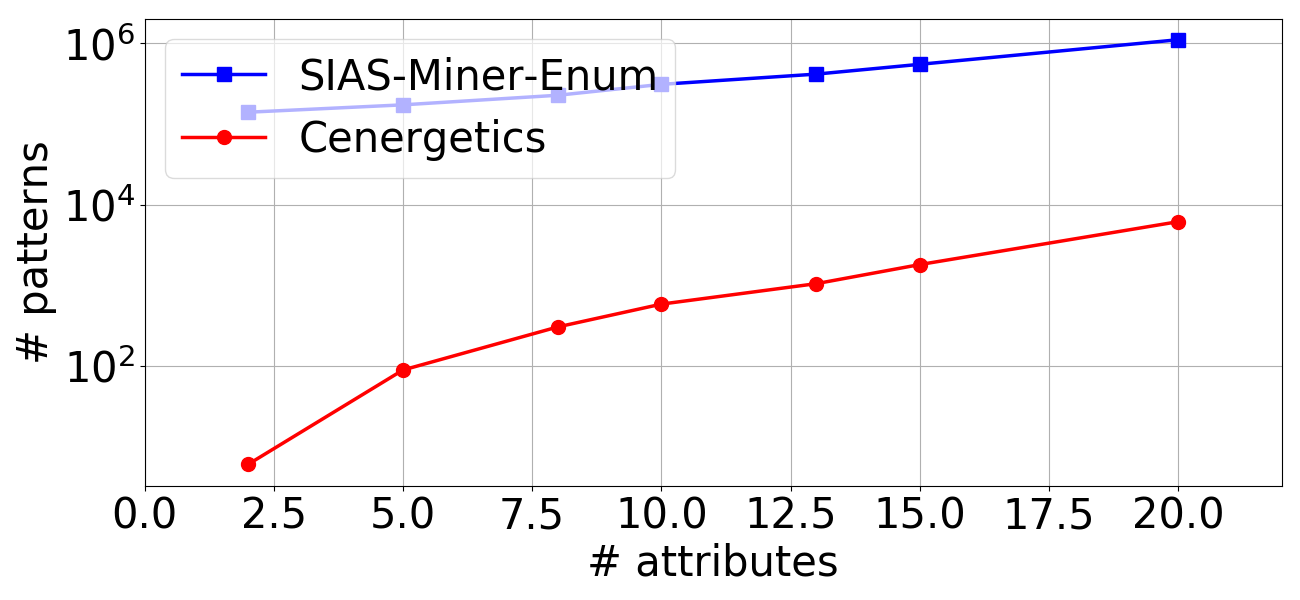} &
    \includegraphics[width=.23\textwidth]{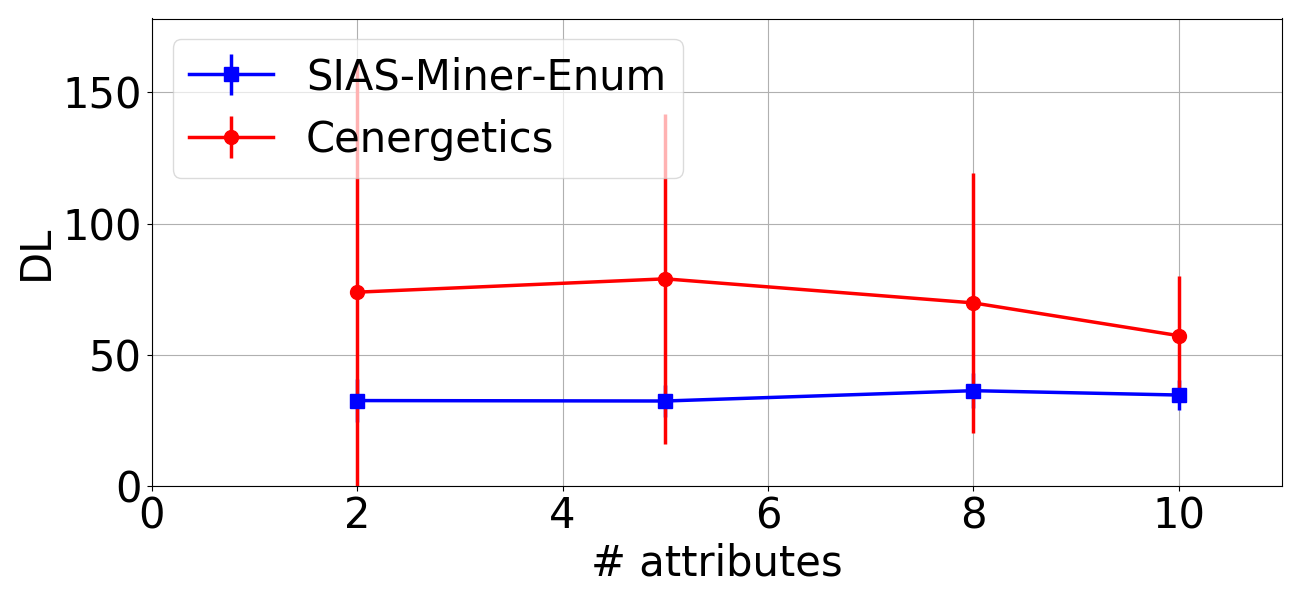}
  &   \includegraphics[width=.23\textwidth]{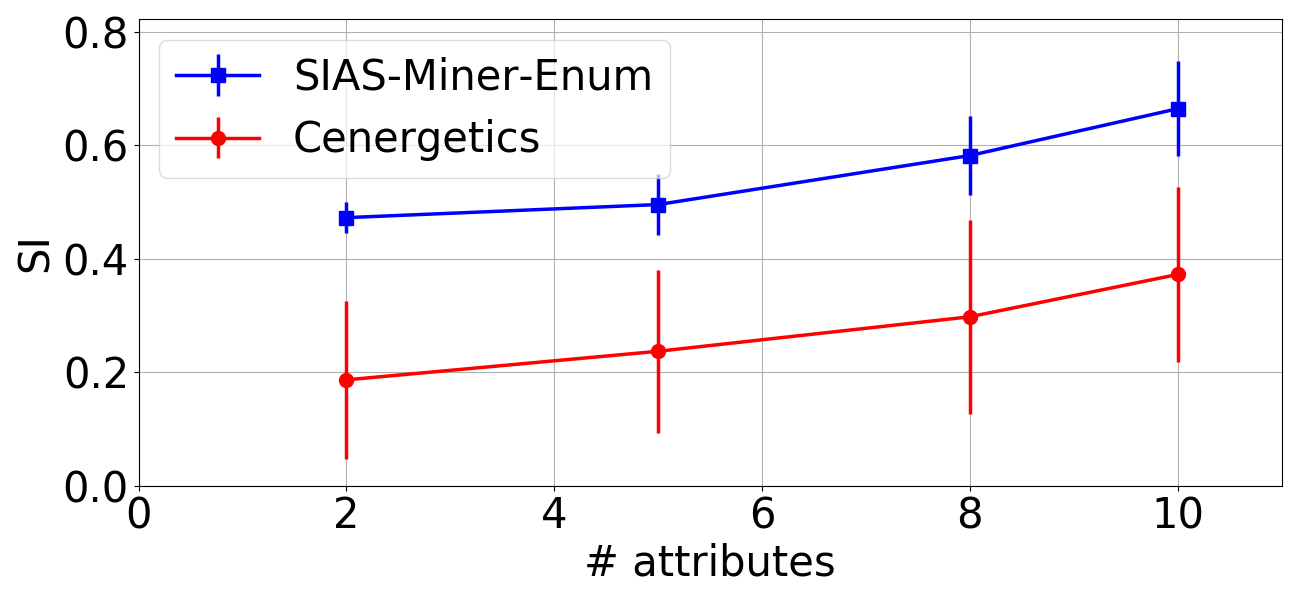}
   \\
  \includegraphics[width=.23\textwidth]{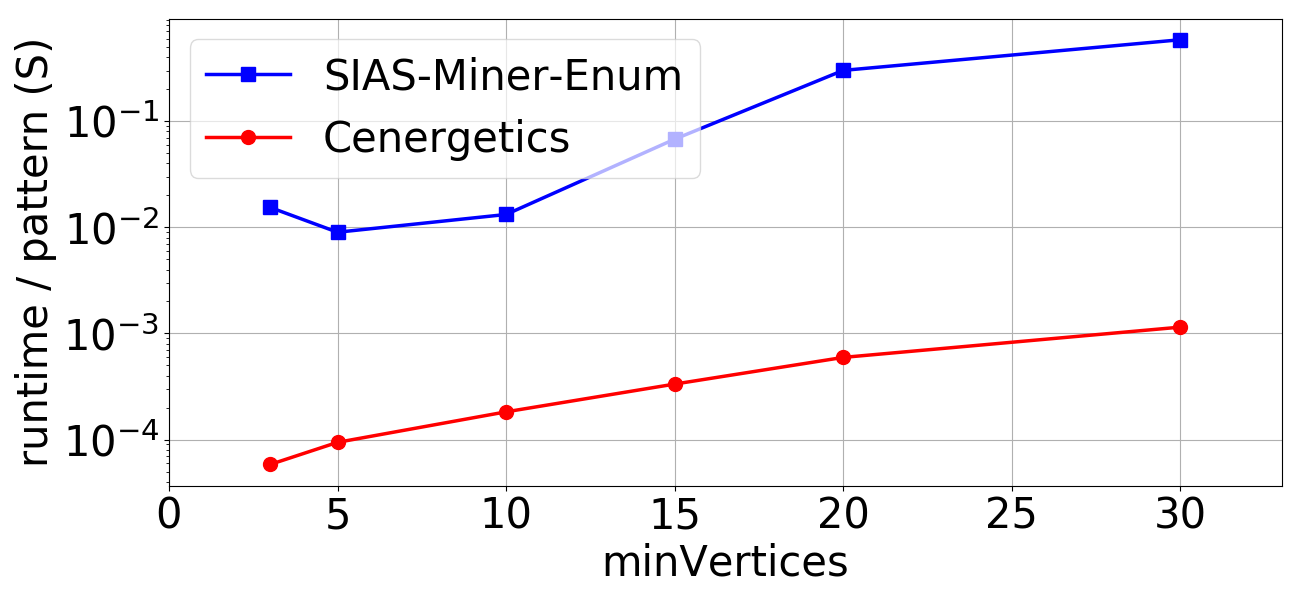} &
   \includegraphics[width=.23\textwidth]{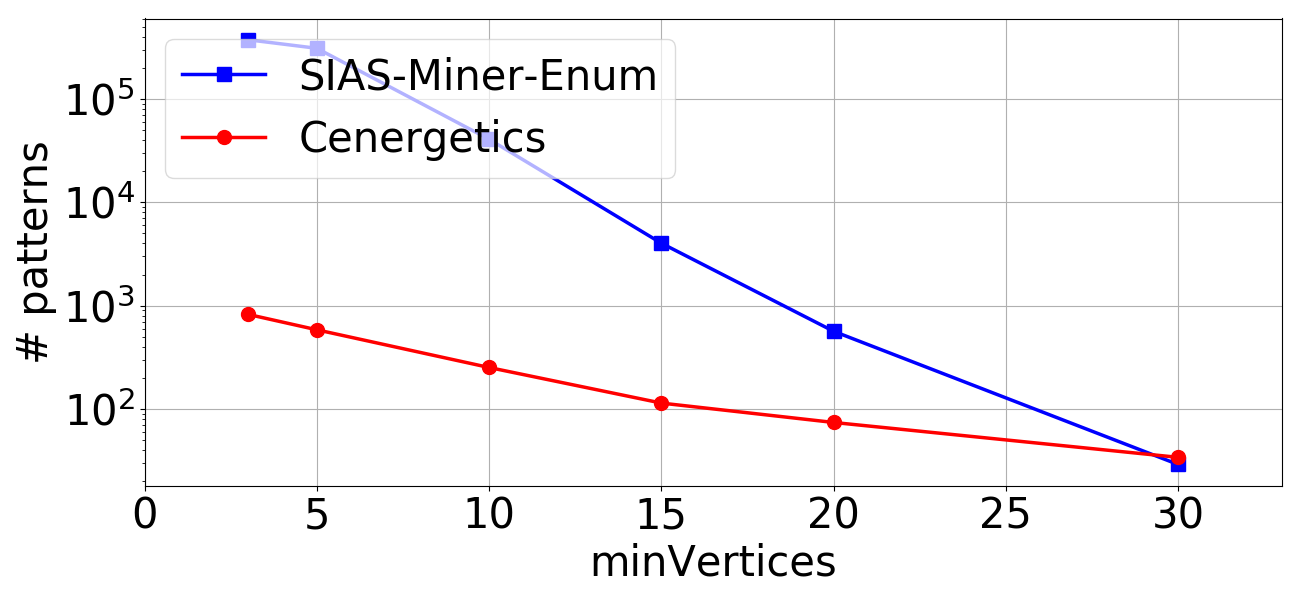} &
    \includegraphics[width=.23\textwidth]{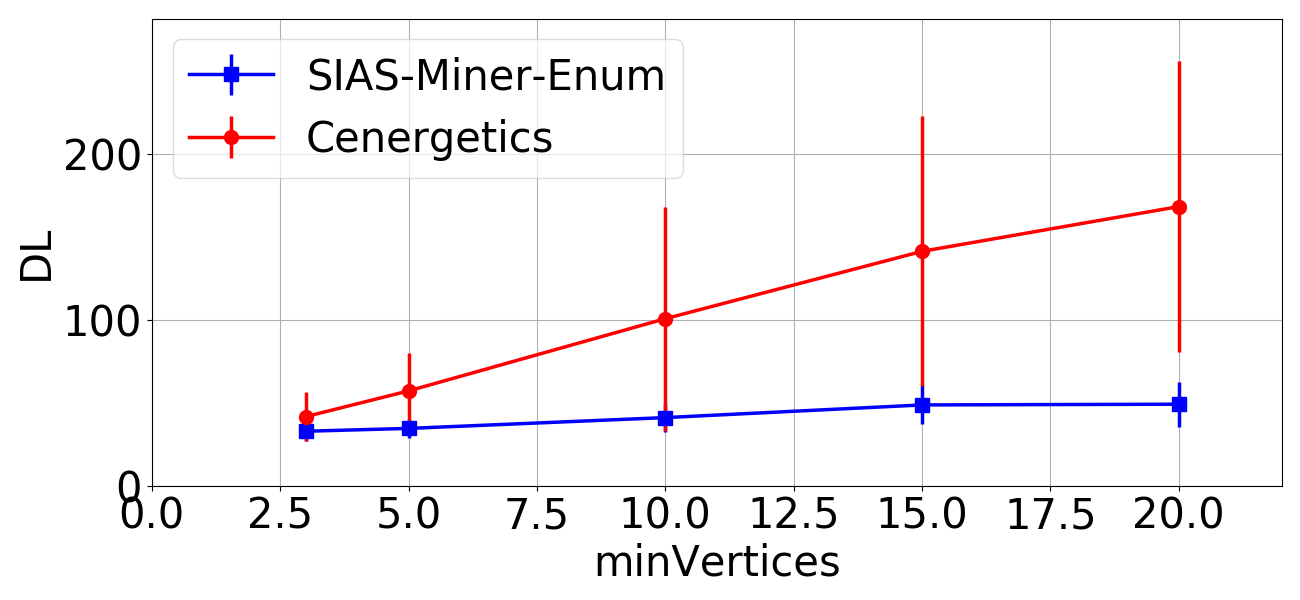} &
     \includegraphics[width=.23\textwidth]{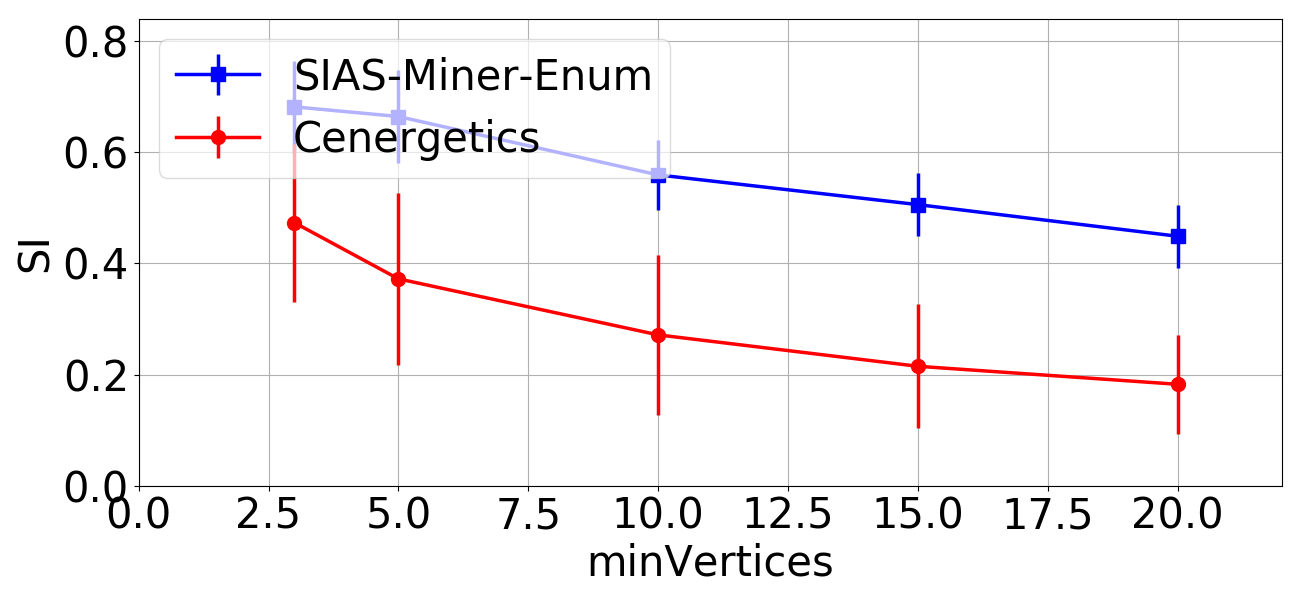}
   \\ 
 \end{tabular}
\caption{\label{fig:quantitativeLondon}\mainAlgo{} vs Cenergetics: runtime per pattern (first column), \#patterns (second column), average description length (third column) and subjective interestingness (fourth column)  of the top $500$ patterns for varying $|V|$ (1st row), $|A|$ (2nd row) and a threshold on the minimum number of
  vertices in searched patterns (3rd row) for London graph ($D=3$).}
\end{figure*}

%
%

\noindent  {\bf Quantitative experiments.}  Fig.~\ref{fig:quantitativeLondon} reports the execution time per pattern, the number of discovered patterns, and the average quality of the top $500$ patterns (i.e., \dl{} and \si{}) of \mainAlgo{} and Cenergetics according to the number of vertices, the number of attributes and the minimum number of vertices of searched patterns. We post-processed the results of Cenergetics in order to obtain similar redescriptions of the vertices as in \mainAlgo{}, but we do not consider this post-processing step in the reported execution times.  These tests reveal that the computational overhead of \mainAlgo{} is important: the discovery of a pattern by \mainAlgo{} is generally one to two orders of magnitude more costly than Cenergetics. However, \mainAlgo{} provides patterns of better quality.  Indeed, the average description length of the top $500$ patterns discovered by  \mainAlgo{} is smaller than those of Cenergetics
and the SI of \csea{} patterns is greater than the one of the patterns extracted by Cenergetics. 

\begin{figure}[t]
\centering
\begin{tabular}{cc}
\includegraphics[width=0.7\columnwidth]{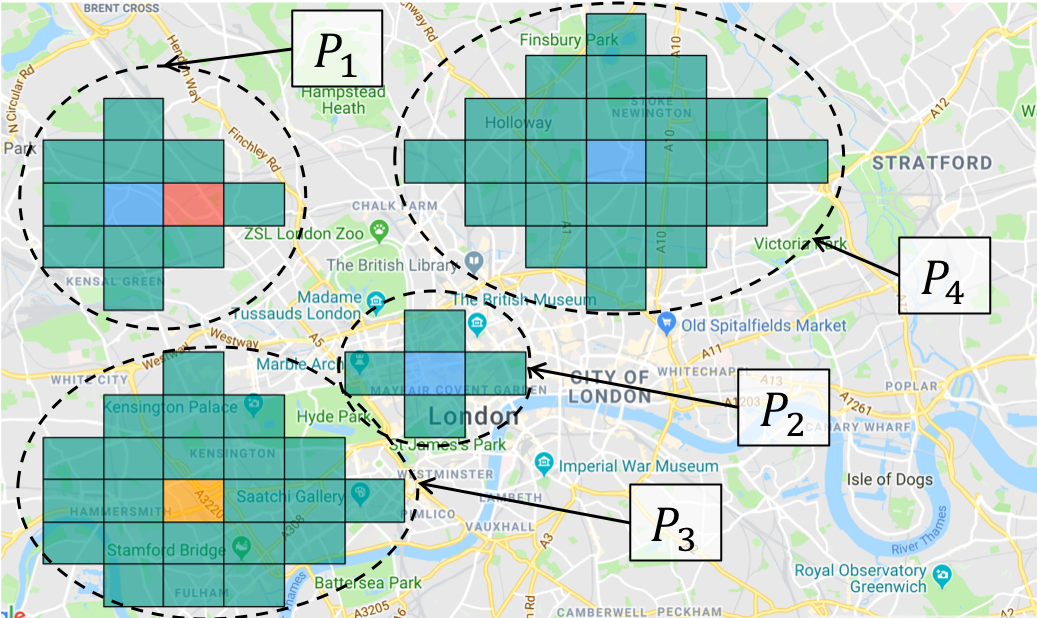}\\
\end{tabular}
\caption{\label{fig:patternsLondonOurAlgo} Top 4 patterns discovered  in London graph by \mainAlgo{}
  ($minVertices=5$, $D=3$). Details are provided in Tab.~\ref{tab:significanceLondon}}
\end{figure}









\begin{figure}[t]
\centering
\begin{tabular}{cc}

\includegraphics[width=0.48\linewidth]{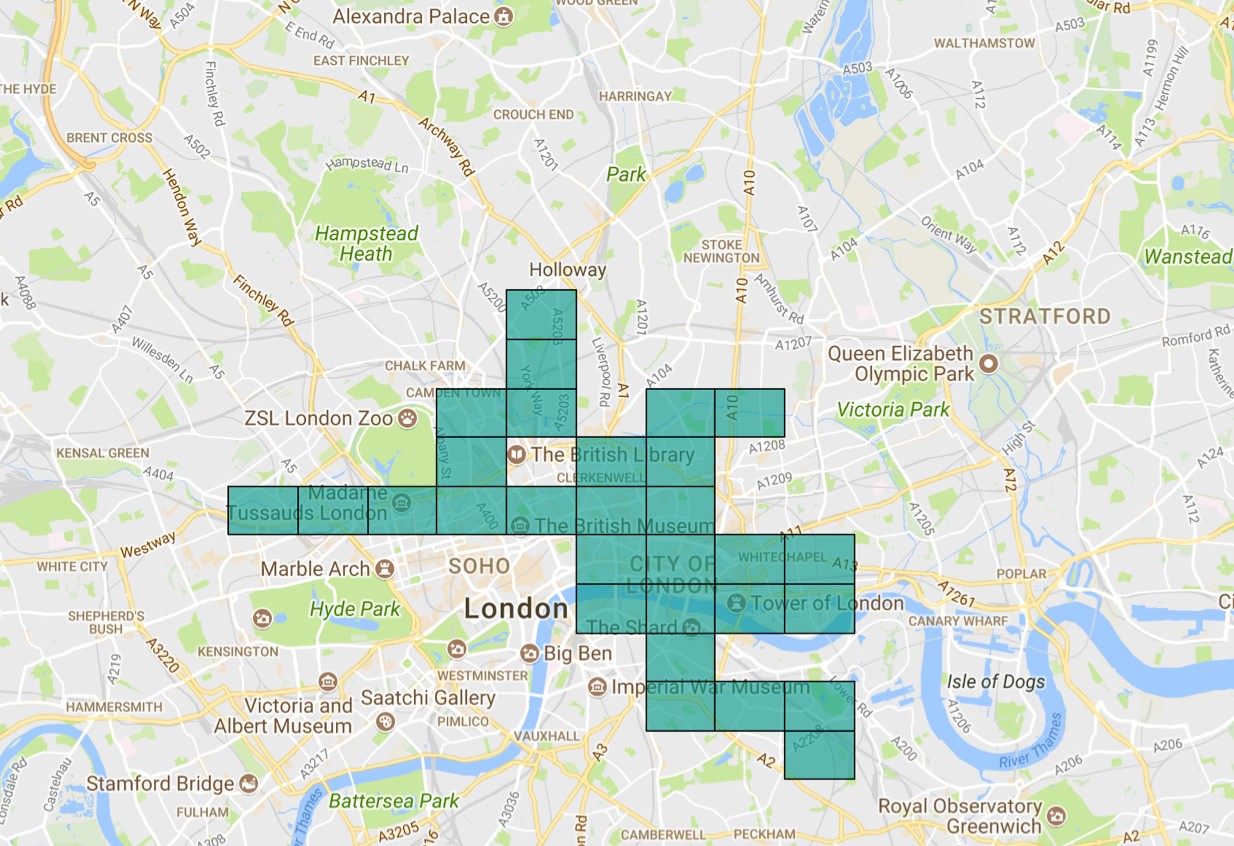} &
\includegraphics[width=0.48\linewidth]{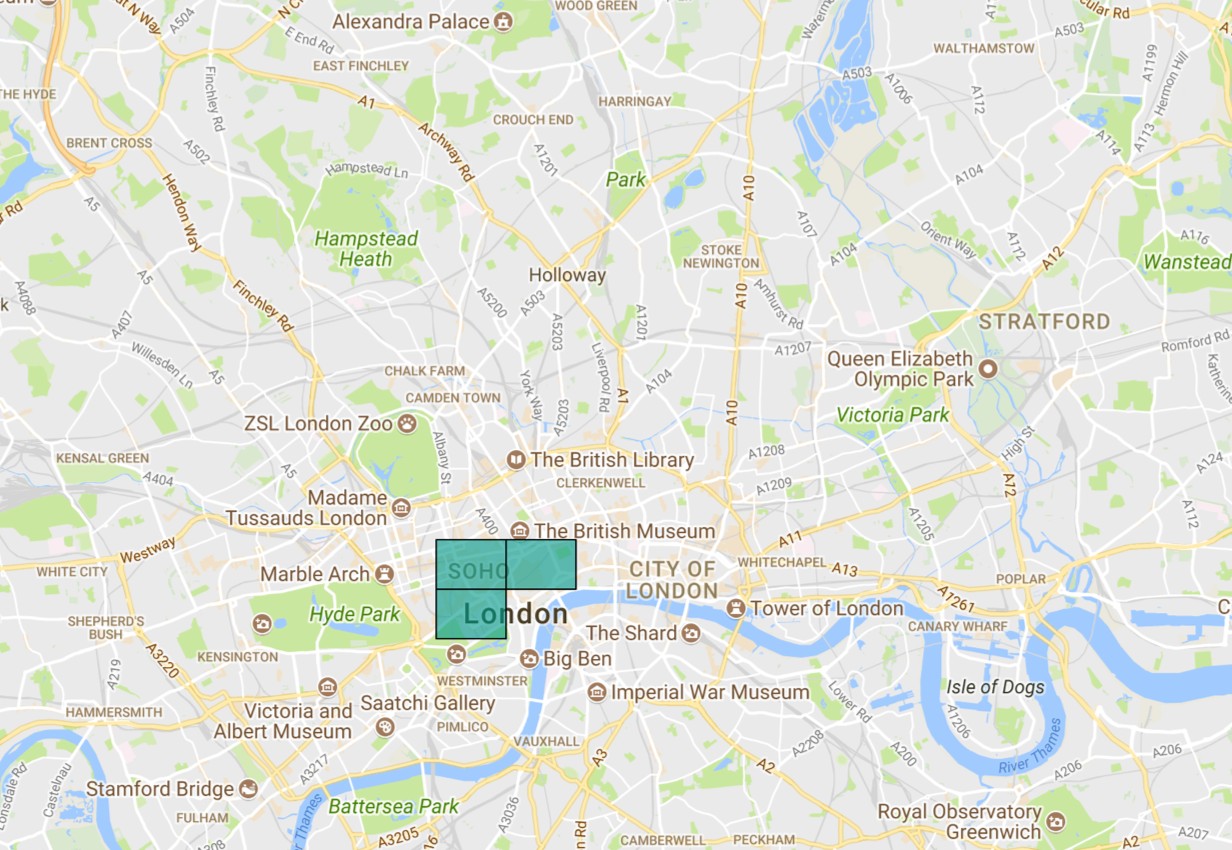} \\

{\scriptsize 
\begin{tabular}{c}
$P_1:$ $\{$professional$\} ^+ $ \\
$\{$shop$\} ^- $ \\
\end{tabular} }
&

{\scriptsize 
\begin{tabular}{c}
$P_2:$ $\{$shop, nightlife, food $\} ^+ $ \\
$\{$outdoors, travel, residence$\} ^- $ \\
\end{tabular} } \\

\includegraphics[width=0.48\linewidth]{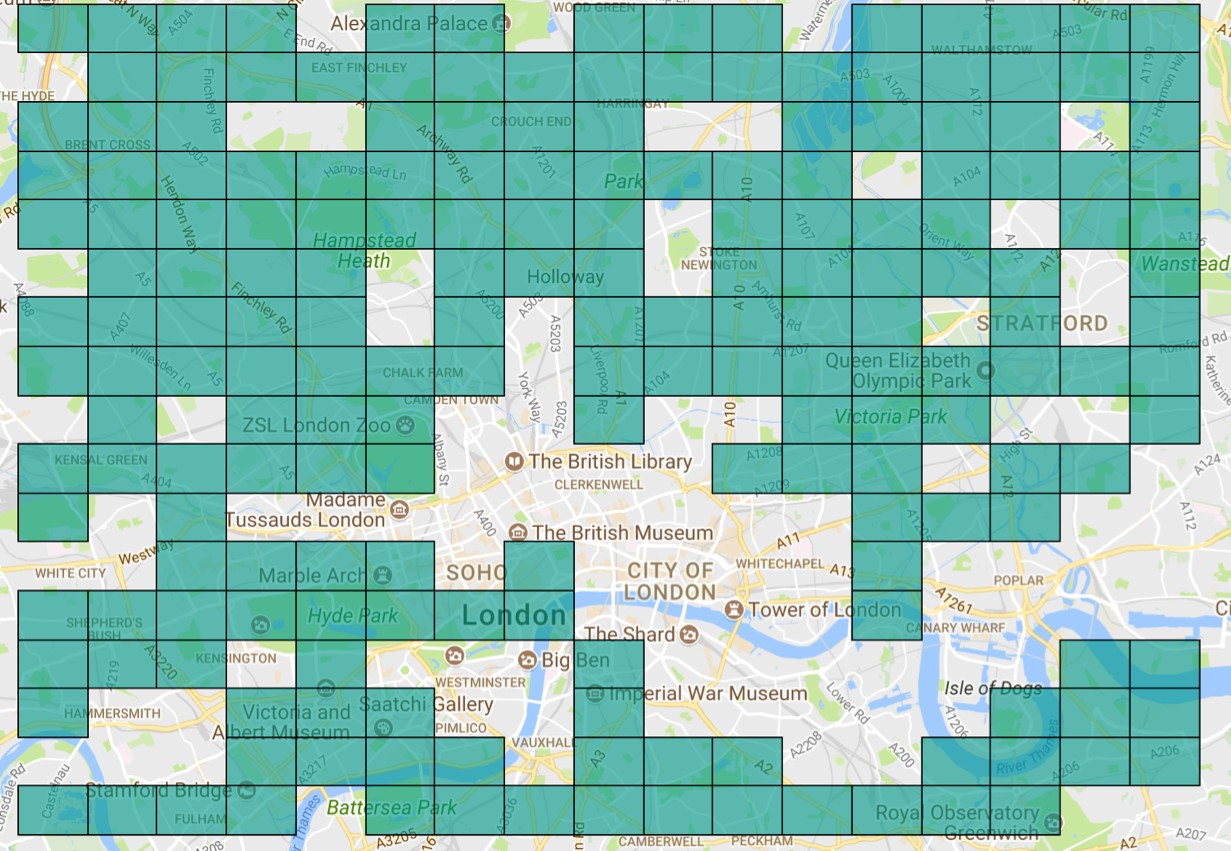}  &
\includegraphics[width=0.48\linewidth]{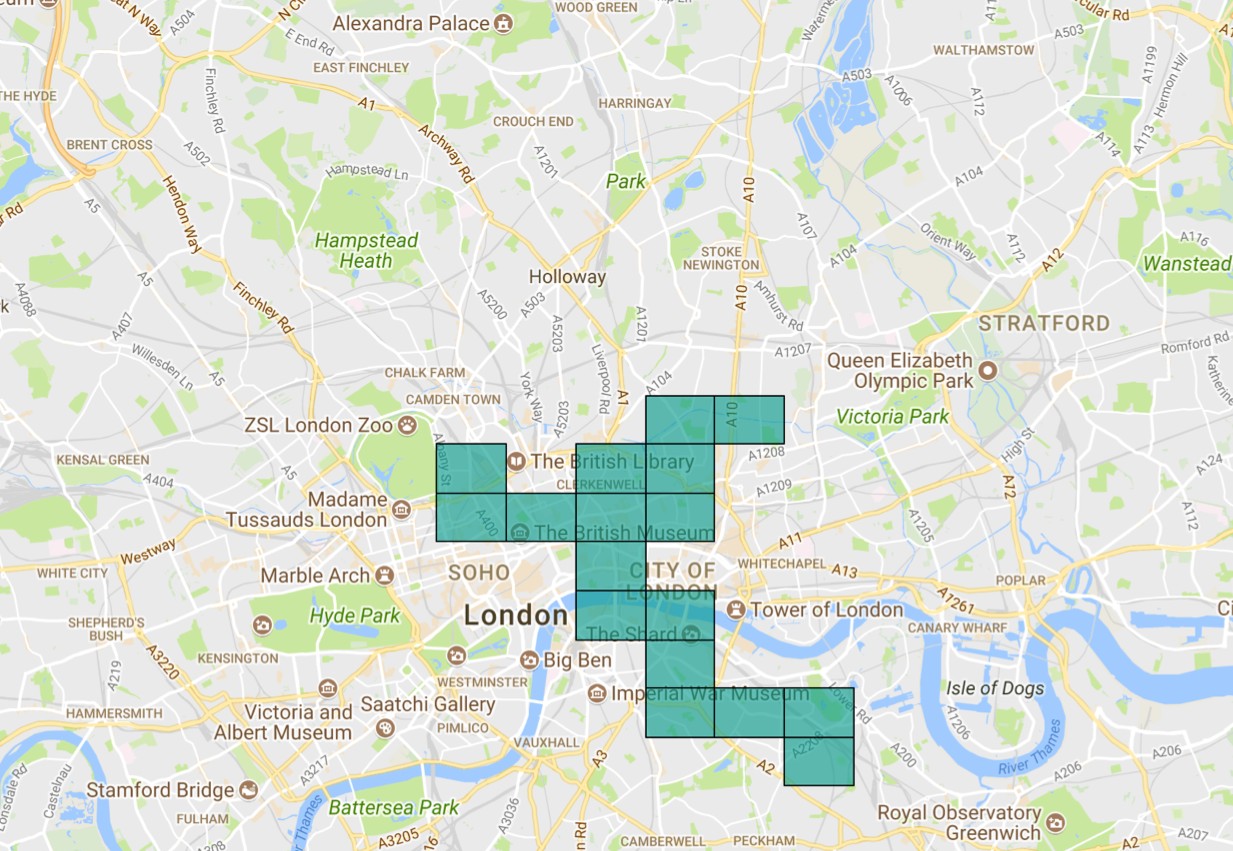} \\

{\scriptsize 
\begin{tabular}{c}
$P_3:$ $\{$professional$\} ^- $ \\
\end{tabular} }

&
{\scriptsize 
\begin{tabular}{c}
$P_4:$ $\{$professional$\} ^+ $, $\{$shop, food$\} ^- $ \\
\end{tabular} }

\\

 


\end{tabular}
\caption{\label{fig:patternsLondonCenergetics} Top 4 patterns discovered  in London graph by Cenergetics
  ($minVertices=5$).}
\end{figure}

\noindent  {\bf Qualitative experiments.} Finally, we show some examples of patterns (with at least 5 vertices) discovered by both \mainAlgo{} and Cenergetics on London graph. The top $4$ \csea{} patterns  discovered by \mainAlgo{} are given in Tab.~\ref{tab:significanceLondon} and displayed in Fig.~\ref{fig:patternsLondonOurAlgo}.  Green cells represent vertices covered by a \csea{} pattern while blue cells  are the centers, purples cells are the centers that do not belong to the pattern, orange cells are centers that are also exception (i.e., behave differently from the pattern but covered by the description)  and the red cells are normal exceptions. 
We also report the top $4$ patterns  discovered by  Cenergetics in Fig.~\ref{fig:patternsLondonCenergetics}. 
Interestingly, \csea{} patterns are more cohesive than Cenergetics ones, described by at most two Neighborhoods, which eases the assimilation by an analyst. 
Unexpectedly, the second best patterns of \mainAlgo{} and Cenergetics are somewhat similar: the \csea{} pattern covers two additional vertices and the two patterns are in agreement on some  overrepresented types of venues  (e.g.,  nightlife, shops). Surprisingly, outdoor venues are consistently overrepresented in the \csea{} pattern while such venues are considered as limited by Cenergetics. This inconsistency may be due to some extreme values in some regions that impact the mean and then the value of the WRacc measure by Cenergetics. 

\begin{table}[t]
\resizebox{\columnwidth}{!}{
   \begin{tabular}{|p{1.5cm}|p{8cm}|}
   \hline
   \textbf{Pattern ID} & \textbf{ Characteristics:} $S=\{(a_i,[l_i,k_i])\}$ \\
   \hline 
   $P_1$  & $\{$food: $[0,0.47]\}^+$ ,  $\{$college: $[1,1]$, event: $[1,1]$, art: $[0.57,1]\}^-$ \\
   \hline
   $P_2$  &   $\{$shop: $[0,0.43]$,  nightlife: $[0,0.44]$,  travel: $[0,0.44]$, college: $[0,0.47]$, outdoors: $[0,0.47]\}^+$ \\
   \hline
      $P_3$  &  $\{$food: $[0,0.31]\}^+$\\
   \hline
      $P_4$  & $\{$food: $[0,0.47]\}^+$ \\
   \hline
   \end{tabular}
   }
   \caption{Detailed characteristics of the top $6$ patterns
     discovered in London dataset by \mainAlgo{} (see Fig.~\ref{fig:patternsLondonOurAlgo}).\label{tab:significanceLondon}}
\end{table}


\noindent  \textbf{Summary.}  Even if our approach has an obvious computational overhead compared to Cenergetics (the problem tackled is  more complex), these experiments show the ability of \mainAlgo{} to discover 
 \csea{} patterns that are more intuitive and informative.











\section{Related work\label{sec:relatedwork}}

Several approaches have been designed to discover new insights in
vertex attributed graphs. The pioneering work of Moser et
al. \cite{DBLP:conf/sdm/MoserCRE09} presents a method to mine dense
homogeneous subgraphs, i.e., subgraphs whose vertices share a large
set of attributes. Similarly, G\"unnemann et al.
\cite{DBLP:conf/icdm/GunnemannFBS10} introduce a method based on
subspace clustering and dense subgraph mining to extract non redundant
subgraphs that are homogeneous with respect to the vertex
attributes.  Silva et al. \cite{DBLP:journals/pvldb/SilvaMZ12} extract pairs made of a dense
subgraph and a Boolean attribute set such that the Boolean attributes
are strongly associated with the dense subgraphs. In
\cite{DBLP:journals/tkde/PradoPRB13}, the authors propose to mine the graph topology of a large attributed
graph by finding regularities among numerical vertex descriptors. 
The main objective of all these approaches is to find regularities instead
of peculiarities within a large graph, whereas \textit{Exceptional
  Subgraph Mining} mines subgraphs with distinguishing characteristics.


Interestingly, a recent work \cite{AtzmuellerDM16} proposes to mine
descriptions of communities from vertex attributes, with a Subgroup
Discovery approach.  In this supervised setting, each community is
treated as a target that can be assessed by well-established measures,
like WRAcc.  In \cite{Kaytoue2017}, the authors aim at discovering contextualized
  subgraphs that are exceptional with respect to a model of the
  data. Restrictions on the attributes, that are associated to
  edges, are used to generate subgraphs. Such patterns are of interest
  if they pass a statistical test and have high value on an adapted
  WRAcc measure. Similarly, \cite{DBLP:conf/kdd/Lemmerich0SHHS16}
  proposes to discover subgroups with exceptional transition behavior
  as assessed by a first-order Markov chain model. The problem of exceptional subgraph mining in attributed graphs was introduced in  \cite{bendimerad2017mining}. 
Based on an adaptation of WRAcc, the method aims to discover subgraphs with homogeneous and exceptional characteristics.   In Section \ref{sec:experiments}, we demonstrate that \csea{} patterns discovered by \mainAlgo{} are more informative and less complex than patterns discovered by the algorithm devised in \cite{bendimerad2017mining}.

 More generally, Subgroup Discovery \cite{DBLP:journals/jmlr/LavracKFT04,DBLP:journals/jmlr/NovakLW09}
aims to find descriptions of sub-populations for which the distribution
of a predefined target value is significantly different from the
distribution in the whole data.  Several quality measures have been defined  to assess the interest of a subgroup. 
The WRAcc is the most commonly used. However, these measures do not take any prior knowledge into account. 
Therefore, we can expect identified subgroups are less informative.  The problem of taking subjective interestingness into account in pattern mining was already identified in \cite{DBLP:conf/kdd/SilberschatzT95} and has seen a renewed interest in the last decade.

The interestingness measure employed here is inspired by the FORSIED framework \cite{DBLP:conf/kdd/Bie11,DBLP:conf/ida/Bie13}, which defines the \si\ of a pattern as the ratio between the \ic\ and the \dl. The \ic\ is the amount of information specified by showing a pattern to the user. The measure is based on the gain from a Maximum Entropy background model that delineates the current knowledge of a user, hence it is \emph{subjective}, i.e., particular to the modeled belief state.

P-N-RMiner \cite{DBLP:journals/ijdsa/LijffijtSKB16}, the tool used here for pattern enumeration, has also been developed under FORSIED. However, the interestingness measure in this paper is very different, because the information contained in the patterns shown to the user does not align with the output of P-N-RMiner. FORSIED has been applied to mine dense subgraphs \cite{van2016subjective}, but not to Exceptional Subgraph Mining, where we also need to account for attribute values. A much faster CP-based implementation of RMiner exists that directly searches for the top-1 most interesting pattern \cite{DBLP:conf/icdm/GunsALB16}. However, this tool does not support structured attributes and the interestingness measure is different, hence it could not be used directly in our problem setting.

\section{Conclusion\label{sec:conclusion}}

We have introduced  a new pattern language in attributed graphs.  A so-called \csea{} pattern provides to the user
a set of attributes  that have exceptional values throughout a subset of vertices.  The strength of the proposed pattern language lies in its independence to a notion of support to assess the interestingness of a pattern. Instead,
the interestingness is defined based on information theory, as the ratio of the information content ($\ic$) over the description length $\dl$. The \ic\ is the amount of information provided by showing the user a pattern. The quantification is based on the gain from a Maximum Entropy background model that delineates the current knowledge of a user. Using a generically applicable prior as background knowledge, we provide a quantification of exceptionality that (subjectively) appears to match our intuition. 
The \dl{} assesses the complexity of reading a pattern, the user being interested in concise and intuitive descriptions. To this end, we proposed to describe a set of vertices as an intersection of neighborhoods within a chosen distane of selected vertices, the distance and vertices making up the description of the subgraph.
We have  shown how an effective and principled algorithm can enumerate patterns of this language. Extensive empirical results on two real-world datasets  confirm that  CSEA patterns are  intuitive, and the interestingness measure aligns well with actual subjective interestingness.
This paper opens up several avenues for further research such as the development of speed-ups of \mainAlgo{} and how to incorporate non-ordinal attribute types in the pattern syntax and interestingness measure.

\noindent{\bf Acknowledgements.} This work was supported by the ERC under the EU's Seventh Framework Programme (FP7/2007-2013) / ERC Grant Agreement no.\ 615517, FWO (project no.\ G091017N, G0F9816N), the EU's Horizon 2020 research and innovation programme and the FWO under the Marie Sklodowska-Curie Grant Agreement no.\ 665501 and  Group Image Mining (GIM) which joins researchers of THALES Group and LIRIS Lab. This work is also partially supported by the CNRS project APRC Conf Pap.

\bibliographystyle{ACM-Reference-Format}
\bibliography{biblio}

\end{document}